\documentclass[draft]{agujournal2019}
\usepackage{url} 
\usepackage{lineno}
\usepackage[inline]{trackchanges} 
\usepackage{soul}
\draftfalse

\journalname{Space Weather}

\begin{document}

%
%

\title{Which Upstream Solar Wind Conditions Matter Most in Predicting $B_z$ within Coronal Mass Ejections}

%
%




\authors{Pete Riley\affil{1}, M. A. Reiss\affil{2,3} and C. M\"ostl\affil{3}}

\affiliation{1}{Predictive Science Inc., San Diego, California, USA.}
\affiliation{2}{Community Coordinated Modeling Center, Code 674, NASA GSFC, Greenbelt, MD 20771, USA}
\affiliation{3}{Austrian Space Weather Office, Zentralanstalt für Meteorologie und Geodynamik, Graz, Austria}






\correspondingauthor{Pete Riley}{pete@predsci.com}




\begin{keypoints}
\item ML algorithms driven by features within the upstream sheath region can accurately predict minimum values of $B_z$ within the ejecta
\item The most important and statistically significant features are sheath number density and total field strength.
\item These features capture compression upstream of the ICME, which correlates with the overall magnetic strength of the ejecta.
\end{keypoints}

%
%

%
%


\begin{abstract}
Accurately predicting the z-component of the interplanetary magnetic field, particularly during the passage of an interplanetary coronal mass ejection (ICME), is a crucial objective for space weather predictions. Currently, only a handful of techniques have been proposed and they remain limited in scope and accuracy. Recently, a robust machine learning (ML) technique was developed for predicting the minimum value of $B_z$ within ICMEs based on a set of 42 `features', that is, variables calculated from measured quantities upstream of the ICME and within its sheath region. In this study, we investigate these so-called explanatory variables in more detail, focusing on those that were (1) statistically significant; and (2) most important. We find that number density and magnetic field strength accounted for a large proportion of the variability. These features capture the degree to which the ICME compresses the ambient solar wind ahead. Intuitively, this makes sense: Energy made available to CMEs as they erupt is partitioned into magnetic and kinetic energy. Thus, more powerful CMEs are launched with larger flux-rope fields (larger $B_z$), at greater speeds, resulting in more sheath compression (increased number density and total field strength).   
\end{abstract}

\section*{Plain Language Summary}

As our society becomes more technologically reliant, the need to accurately forecast the severity of geomagnetic storms becomes increasingly important. Storms driven by fast coronal mass ejections (CMEs) represent the biggest threat, being responsible for all of the major geomagnetic events in recorded history, in part, because they contain the largest magnetic fields within them. Fast CMEs ploughing through the solar wind produce a so-called ``sheath'' region ahead of them. In this study, we relate the properties of the sheath region to the size of the CME's magnetic field, demonstrating that estimates of compression provide the most robust predictions of the ensuing magnetic field. This relationship is a promising forecasting approach that could provide more than one day's advance warning before the arrival of the peak magnetic fields within the CME. 

%
%

%


%
%
%
%

\section{Introduction}

An accurate prediction of the large-scale interplanetary magnetic field, and, in particular, its z-component, $B_z$ is a crucial parameter for any space weather forecasting system, and yet, thus far, it has remained largely illusive (a point exemplified by the fact that no prediction centre currently provides a forecast for $B_z$). Many processes contribute to a non-zero z-component of the IMF. However, in the absence of transient effects, the large-scale quiescent spiral heliospheric magnetic field has no net $B_z$. Waves and turbulence can be superposed on top of this large-scale picture 
\cite<e.g.>[]{horbury01a}, but, in and of themselves, these fluctuations do not actively drive substantial space weather.  From a geo-effective viewpoint, large solar eruptions generating coherent flux rope structures that propagate relatively undisturbed to 1 au represent the major source of geomagnetic storms, particularly if the axis of the flux rope lies in, or near to, the ecliptic plane. In addition, fast CMEs drive fast-mode shocks ahead of them that compress the IMF, amplifying the wave/turbulent fluctuations. Furthermore, draping of the large-scale field around the ejecta can result in large, sustained values of $B_z$ \cite{gosling87c}. Thus, from a prioritised, operational perspective, it is the fast, coherent CMEs, driving a strong sheath region that require our attention \cite{lugaz16a}.  

Several promising approaches for predicting $B_z$ have been developed over the years, ranging from strictly statistical techniques (e.g., \citeA{riley17b,chen96b}) to self-consistent, physics-based global magnetohydrodynamic (MHD) simulations (e.g., \citeA{riley07a,shiota16a,torok18a}). Between these extremes is a range of empirically-based approaches that combine observations with limited empirical or statistical modelling to optimize predictive capabilities (e.g., \citeA{savani15a,mostl17a}).  

Machine learning (ML) techniques have become widely adopted within many scientific fields (e.g., \citeA{witten05a}). Solar and heliospheric research, in particular, is currently benefitting from the application of both supervised and unsupervised techniques to understand better, or at least describe and predict, a wide array of phenomena (e.g., \citeA{qahwaji08a,bobra14a,camporeale17a, camporeale18a,heidrich18a,bailey20a,camporeale20a}.  Supervised ML algorithms, in particular, can be effective approaches for solving regression problems. More straightforward multiple regression approaches remain valid; however, more complex techniques, such as the random forest algorithm, promise to mitigate issues such as over-fitting, hyper-parameter tuning, and handling missing data (e.g., \citeA{xia17a}). An important caveat, however, is that while simpler algorithms, such as simple multiple regression, can be easily understood, self-coded, or at least carefully examined, many of the ML algorithms must be treated as ``black boxes.'' The use of standard packages containing the routines, which have been tested extensively across many scientific domains, provides reassurance that the implementation of the algorithms is correct and robust.  

In a previous study, we applied ML methodology to assess whether upstream in-situ ICME sheath region measurements could provide estimates of the resulting (1) minimum value in the $B_z$ component of the magnetic field; or (2) maximum value of the total field ($|B|)$, within the following ICME~\cite{reiss21a}. We developed a predictive model based on 348 ICMEs observed by Wind, STEREO-A, and STEREO-B spacecraft. We found relatively high associations (Pearson Correlation Coefficient, PCC = 0.71 and 0.91, respectively) between the target variable and a set of 42 input (explanatory) variables or features. These explanatory variables were made up of various statistical properties of the three components of the magnetic field, total magnetic field strength, plasma density, temperature, and bulk solar wind speed, specifically: the mean value, standard deviation, minimum and maximum values, the ratio between the maximum and minimum values, and the ratio between the mean value and standard deviation (i.e., the coefficient of variation). Thus, for each of the seven variables, there were six statistical measures, resulting in 42 features or input variables that could, in principle, explain the observed variations in either the minimum value of $B_z$ within the ejecta, or the maximum value of the field. Since the objective was to derive a predictive model, we separated the 348 events into a training (4/5) and evaluation (1/5) dataset and applied three models (linear regressor (LR), random forest regressor (RFR), and gradient boosting regressor (GBR)). The high resulting PCC values suggested that this might be a promising forecast tool for estimating the strength of the ICME's magnetic field many hours before its arrival at 1 au. The original study, however, left several questions unanswered. First, which of the 42 input variables were most important, that is, which variables were responsible for most variations in $B_z$ within the ICME? Second, which variables were statistically significant? Third, scientifically, what was the underlying mechanism for the strong association between at least some of the input variables and the target variables? 

In this study, we aim to build on this previous study by addressing these questions. Specifically, we seek to identify those variables that are statistically significant and those that contribute most to the variability in the targets and to provide an explanation for why. In doing so, we believe that this also addresses a concern often raised with respect to the application of ML techniques within Heliophysics: What do we learn from their use? While the previous study \cite{reiss21a} directly addressed an operational space weather need and, thus, was not primarily concerned with these issues, the present study aims to use ML approaches to better understand why such high correlations are present. 
 
In the Sections that follow, we first introduce the dataset and outline how it is pre-processed, then summarize the ML techniques we plan to apply. We next describe one set of analyses in detail and provide summaries of several re-analyses based on different initial processing of the data. Finally, we interpret the results from an intuitive picture of ICME evolution, discuss the limitations of this study, and provide suggestions for how this study could be improved upon in the future. 

\section{Methodology}

\subsection{Data}

\citeA{reiss21a} identified 364 ICMEs that produced an upstream sheath region from the HELIO4CAST ICME catalogue \cite{mostl20a}. The requirement that they drive a sheath ahead (or, technically, displayed a density enhancement) likely meant that these ICMEs travelled faster than the ambient wind within which they were embedded, although the compression could be produced by expansion of a slower ejecta \cite{siscoe08a,salman21a}. All events were observed by Earth-based or STEREO-A/B spacecraft between January 2007 and March 2021, and thus, were all located at approximately 1 au. The list was further pruned to 348 events for which there was good data coverage. 

Additionally, although the original catalogue did not contain any information about whether or not a shock preceded the CME, we cross-referenced the HELIO4CAST ICME catalogue with an independent catalogue of CME-driven shocks identified in the Wind dataset \cite{nieves18a}, and created a new dataset of events with a 0/1 indicating whether a shock was, or was not present. Of the total number of events (348), 247 were found to be driving an interplanetary shock, while for 101 events, no obviously associated shock could be found. 

From the time series of these events, six statistical measures (mean, standard deviation (std), minimum (min), maximum (max), the ratio of maximum to minimum values (minmax), and the coefficient of variation (cv), i.e., the ratio of the mean to standard deviation) were calculated for each of the following variables: Total magnetic field ($B_t$), three components of the field ($B_x$, $B_y$, $B_z$), bulk plasma speed ($v_t$), number density ($n_p$), and proton temperature ($T_p$). The vector quantities are given in the Heliocentric Earth equatorial (HEEQ) system, based on the Sun's rotation axis, where the z-axis is parallel to the Sun's rotation axis (positive northward), the x-axis points towards the intersection of the solar equator and the solar central meridian as seen from Earth, and the y-axis completes the right-hand orthogonal system. 
The temporal boundaries used to compute these quantities were from the start of the sheath to the end of the sheath interval. \citeA{reiss21a} also investigated the effects of adding an additional four hours of data at the start of each window, thus, incorporating a small portion of the actual ejecta into the explanatory variables. 

The dataset (containing $348 \times 42$ (14616) points) is summarised in Table~\ref{tab1}, which shows each feature in column 1. For each, the mean, standard deviation, minimum, 25th and 75th percentiles, and maximum values are given. Note that these are the statistics of the features, which are also statistical measures themselves. Thus, the `St. Dev.' (column 4) of `std.bx' is the standard deviation of the standard deviations computed for each of the 348 sheath regions. The last row summarizes the `Target' or output variable, which in this case is the minimum value of $B_z$ within the ejecta. The abbreviation `std' refers to the standard deviation, `cv' is the coefficient of variation, or the ratio between the mean value and the standard deviation, and `minmax' is the ratio between the maximum and minimum values. For convenience, $bt$ refers to the total magnetic field component, $|B|$, while the other variables are easily understood. 

\begin{sidewaystable} \centering 
\scriptsize
  \caption{List of parameters used in the multiple-regression analysis, together with their basic statistical properties. See the text for a detailed explanation of each parameter.} 
\begin{tabular}{@{\extracolsep{5pt}}lccccccc} 
\\[-1.8ex]\hline 
\hline \\[-1.8ex] 
Feature & \multicolumn{1}{c}{N} & \multicolumn{1}{c}{Mean} & \multicolumn{1}{c}{St. Dev.} & \multicolumn{1}{c}{Min} & \multicolumn{1}{c}{Pctl(25)} & \multicolumn{1}{c}{Pctl(75)} & \multicolumn{1}{c}{Max} \\ 
\hline \\[-1.8ex] 
mean.bx. & 348 & $-$0.292 & 2.737 & $-$8.388 & $-$2.093 & 1.383 & 11.411 \\ 
max.bx. & 348 & 7.114 & 4.736 & $-$4.941 & 3.838 & 9.357 & 32.120 \\ 
std.bx. & 348 & 3.239 & 1.914 & 0.000 & 1.946 & 4.206 & 11.714 \\ 
min.bx. & 348 & $-$7.522 & 4.609 & $-$27.209 & $-$10.101 & $-$4.387 & 3.577 \\ 
cv.bx. & 348 & 12.911 & 130.146 & 0.000 & 0.843 & 3.735 & 2,416.322 \\ 
minmax.bx. & 348 & $-$1.745 & 15.220 & $-$265.063 & $-$1.506 & $-$0.692 & 78.281 \\ 
mean.by. & 348 & 0.472 & 4.909 & $-$21.306 & $-$2.847 & 3.688 & 16.800 \\ 
max.by. & 348 & 10.184 & 6.908 & $-$10.915 & 6.109 & 13.074 & 43.805 \\ 
std.by. & 348 & 4.591 & 3.270 & 0.000 & 2.536 & 5.908 & 25.036 \\ 
min.by. & 348 & $-$9.401 & 7.095 & $-$50.495 & $-$12.904 & $-$4.934 & 5.030 \\ 
cv.by. & 348 & 3.519 & 8.915 & 0.000 & 0.567 & 2.633 & 107.937 \\ 
minmax.by. & 348 & $-$1.049 & 9.981 & $-$138.319 & $-$1.277 & $-$0.535 & 120.411 \\ 
mean.bz. & 348 & 0.229 & 3.109 & $-$9.976 & $-$1.591 & 1.966 & 13.673 \\ 
max.bz. & 348 & 10.256 & 6.910 & $-$5.495 & 5.821 & 13.033 & 60.193 \\ 
std.bz. & 348 & 4.558 & 2.968 & 0.000 & 2.667 & 5.790 & 20.432 \\ 
min.bz. & 348 & $-$9.799 & 6.736 & $-$40.842 & $-$12.236 & $-$5.743 & 7.602 \\ 
cv.bz. & 348 & 19.078 & 213.734 & 0.000 & 1.251 & 5.463 & 3,962.469 \\ 
minmax.bz. & 348 & $-$0.957 & 2.311 & $-$22.105 & $-$1.209 & $-$0.710 & 32.596 \\ 
mean.bt. & 348 & 9.399 & 4.546 & 2.026 & 6.417 & 11.183 & 28.042 \\ 
max.bt. & 348 & 14.362 & 7.992 & 2.681 & 8.777 & 18.131 & 61.447 \\ 
std.bt. & 348 & 2.253 & 1.741 & 0.000 & 1.122 & 2.901 & 14.293 \\ 
min.bt. & 348 & 3.076 & 2.274 & 0.228 & 1.535 & 4.115 & 16.107 \\ 
cv.bt. & 348 & 0.230 & 0.105 & 0.000 & 0.155 & 0.284 & 0.734 \\ 
minmax.bt. & 348 & 0.255 & 0.188 & 0.014 & 0.105 & 0.355 & 1.000 \\ 
mean.vt. & 348 & 447.167 & 106.012 & 283.544 & 369.984 & 494.505 & 823.933 \\ 
max.vt. & 348 & 498.517 & 131.658 & 300.348 & 401.584 & 567.684 & 1,001.504 \\ 
std.vt. & 348 & 19.926 & 15.332 & 0.000 & 8.860 & 27.290 & 119.767 \\ 
min.vt. & 348 & 400.165 & 88.781 & 239.981 & 337.344 & 439.992 & 723.750 \\ 
cv.vt. & 348 & 0.043 & 0.028 & 0.000 & 0.023 & 0.055 & 0.208 \\ 
minmax.vt. & 348 & 0.816 & 0.092 & 0.408 & 0.756 & 0.885 & 1.000 \\ 
mean.np. & 348 & 14.102 & 9.264 & 0.451 & 7.685 & 18.491 & 84.160 \\ 
max.np. & 348 & 28.930 & 20.587 & 1.164 & 14.502 & 37.137 & 147.326 \\ 
std.np. & 348 & 4.704 & 4.021 & 0.000 & 1.970 & 6.105 & 26.247 \\ 
min.np. & 348 & 5.812 & 4.904 & 0.112 & 2.728 & 7.271 & 49.334 \\ 
cv.np. & 348 & 0.326 & 0.173 & 0.000 & 0.201 & 0.422 & 1.135 \\ 
minmax.np. & 348 & 0.248 & 0.166 & 0.002 & 0.127 & 0.343 & 1.000 \\ 
mean.tp. & 348 & 153,878.500 & 150,688.400 & 13,725.420 & 52,988.160 & 207,266.300 & 964,935.600 \\ 
max.tp. & 348 & 560,333.500 & 751,663.600 & 19,754.490 & 117,018.800 & 647,930.000 & 5,249,402.000 \\ 
std.tp. & 348 & 74,633.040 & 96,655.200 & 0.000 & 15,796.020 & 91,670.020 & 659,472.700 \\ 
min.tp. & 348 & 45,772.240 & 44,346.380 & 4,565.096 & 20,026.440 & 54,835.390 & 360,332.500 \\ 
cv.tp. & 348 & 0.414 & 0.207 & 0.000 & 0.270 & 0.522 & 1.548 \\ 
minmax.tp. & 348 & 0.167 & 0.136 & 0.004 & 0.066 & 0.248 & 1.000 \\ 
Target & 348 & $-$9.456 & 6.681 & $-$60.782 & $-$11.760 & $-$5.506 & 0.000 \\ 
\hline \\[-1.8ex] 
\end{tabular}  
\label{tab1}
\end{sidewaystable} 

Although we will refer back to these values later, it is worth noting that they are, in some cases, quite different from the average properties of the ambient solar wind. For example, the mean number density (mean.np.) for the sheath regions is 14.1~cm$^{-3}$, which is more than twice the average number density for the background solar wind at 1 au. Moreover, the maximum value (84.2 cm$^{-3}$) is more than an order of magnitude higher. 

To assess the effects of varying data ranges on the model results, these data were re-normalized in one of several ways. This can mitigate potential biases from some features having a much broader range and, hence, a disproportionate effect on the results. First, a min-max procedure was applied where all variables were re-scaled such that the minimum value was set to zero and the maximum value to one. While this potentially levels all data to the same range, if outliers are present, it can have the reverse effect of squeezing some datasets to a disproportionately smaller range. Thus, a second approach of z-score re-normalisation was also applied. This transformation refers to the process of re-normalizing each value in a dataset so that the mean of all of the values is 0 and the standard deviation is 1. The effects of wide or narrow data ranges are addressed as well as mitigating the impacts of outliers. The only minor impact lies in visual interpretation, where the min-max procedure is advantageous in allowing all data to be viewed between zero and one. With the exception of Table~\ref{tab1}, all statistical model results are reported in relation to z-score re-normalization.  

Since the original study was completed \cite{reiss21a}, an update to the HELIO4CAST ICME catalogue has been published online (ICMECATv2.1
). This contains 521 events with sheath regions from Wind 1995-2021, and STEREO-A/B 2007 - 2021. To more directly link the current study with the original, we primarily analyse events from the original dataset. We repeated the analysis with the v2.1 data, however, to verify that the main inferences and conclusions drawn did not change with the addition of more events.  

\subsection{Models}

In this study, multiple regression models were analyzed using several packages in $R$, primarily relying on the ``Fitting Linear Models'' suite (lm) \cite{r20a}. These regression techniques, however, only identify the statistical significance of the variables. To estimate the relative importance of each explanatory variable in describing the variability in depth, we applied traditional statistical and more modern machine learning (ML) approaches.

To identify the explanatory variables that provided the best-performing model (i.e., one that lowered the prediction error), we used the 'stepAIC' function from the MASS package \cite{mass20a}. Both forward selection and backward selection (i.e., backward elimination) were performed \cite{james13a}. In the former, initially, no predictors are included in the model, and the algorithm iteratively adds the most contributory variables, stopping when the improvement is no longer statistically significant. 

Using multiple approaches is important for assessing the uncertainty that should be ascribed to a particular ordering of the variables since different techniques rely on different metrics for importance. The techniques applied included: random forest \cite{liaw02a}, Xgboost \cite{chen16a}, relative importance \cite{gromping06a}, MARS via earth \cite{earth19a}, step-wise regression \cite{bendel76a}, and DALEX \cite{biecek18a}. These were chosen because they are representative of the most widely applied methods. 

It is important to reiterate that these techniques use different definitions of what signifies ``important'', and, thus, we do not expect complete agreement amongst the results. Nevertheless, where the results do agree is where we can be most confident, and where they do not, we must remain more cautious.

It is worth remarking that we chose to use a broad range of ML techniques to highlight both the robustness of the approaches, but also their limitations. Thus, together, they could be thought of as an ``ensemble'' of models where agreements can be used to increase our confidence and disagreements to reduce it. On the other hand, it is beyond the scope of the present study to attempt any kind of ensemble forecast framework using realizations of the different approaches. 

We also note that our purpose here is to apply these statistical models to understand which of these explanatory variables are most important for explaining the variations in $B_z$ within the ejecta. In particular, this was not a predictive exercise, as was undertaken by \citeA{reiss21a}; our aim is to understand why the ML techniques perform so well and what they teach us about what is driving these associations.   

\section{Results}

When a multiple linear regression analysis was performed on the 42-parameter dataset, five explanatory variables were found to be most significant ($p < 0.05$): mean.bt., std.bt., cv.bt., mean.np., and max.np. (Table~\ref{tab2}). A low p-value indicates that we can reject the null hypothesis that the relation between the explanatory variable and the target ($B_z$ within the ICME) is statistically insignificant. Together, the goodness-of-fit measure $R^2$ was 0.579, and thus, these variables, together, are capable of explaining 58\% of the variations in $B_z$ within the ICME. 

Focusing specifically on these variables, we repeated the regression analysis. This led to the results shown in Table~\ref{tab3}. From this, we infer that five of the variables are statistically significant at the $p < 0.01$ level (mean.bt., std.bt., cv.bt.,  mean.np., and max.np.). In general, we infer that the parameters associated with $|B|$ and $n_p$ appear to be statistically significant. Intuitively, this makes sense that $|B|$ and $n_p$ would be correlated within compression regions.  

\begin{table}[!ht] \centering 
\scriptsize
  \caption{Summary of multiple-regression analysis on 42 explanatory variables described in Table 1. For brevity, some of the entries that were found not to be statistically significant have been omitted. The terms in parentheses represent the standard error (an estimate of the standard deviation of the coefficient), while the Residual Std. Error measures how well a regression model fits a dataset, and the F Statistic tests the significance of regression coefficients in linear regression models.} 
  \label{tab2} 
\begin{tabular}{@{\extracolsep{5pt}}lc} 
\\[-1.8ex]\hline 
\hline \\[-1.8ex] 
 & \multicolumn{1}{c}{\textit{Dependent variable:}} \\ 
\cline{2-2} 
\\[-1.8ex] & Target \\ 
\hline \\[-1.8ex] 
 max.bx. & 0.047 \\ 
  & (0.114) \\ 
 std.bx. & 0.136 \\ 
  & (0.131) \\ 
 min.bx. & 0.056 \\ 
  & (0.111) \\ 
 cv.bx. & $-$0.064 \\ 
  & (0.039) \\ 
 minmax.bx. & $-$0.045 \\ 
  & (0.038) \\ 
 mean.by. & 0.011 \\ 
  & (0.077) \\ 
 max.by. & 0.092 \\ 
  & (0.136) \\ 
 std.by. & $-$0.200 \\ 
  & (0.141) \\ 
 min.by. & 0.118 \\ 
  & (0.151) \\ 
 cv.by. & 0.011 \\ 
  & (0.041) \\ 
 minmax.by. & 0.036 \\ 
  & (0.038) \\ 
... \\
mean.bt. & 0.591$^{***}$ \\ 
  & (0.196) \\ 
 max.bt. & $-$0.481 \\ 
  & (0.295) \\ 
 std.bt. & $-$0.775$^{***}$ \\ 
  & (0.260) \\ 
 min.bt. & $-$0.083 \\ 
  & (0.132) \\ 
 cv.bt. & 0.648$^{***}$ \\ 
  & (0.141) \\ 
...\\
 mean.np. & $-$0.689$^{***}$ \\ 
  & (0.194) \\ 
 max.np. & 0.440$^{**}$ \\ 
  & (0.182) \\ 
 std.np. & 0.012 \\ 
  & (0.192) \\ 
 min.np. & $-$0.083 \\ 
  & (0.131) \\ 
 cv.np. & $-$0.183$^{*}$ \\ 
  & (0.111) \\ 
 ... \\
 minmax.tp. & 0.148$^{*}$ \\ 
  & (0.083) \\ 
 Constant & $-$0.000 \\ 
  & (0.037) \\ 
\hline \\[-1.8ex] 
Observations & 348 \\ 
R$^{2}$ & 0.579 \\ 
Adjusted R$^{2}$ & 0.523 \\ 
Residual Std. Error & 0.691 (df = 306) \\ 
F Statistic & 10.280$^{***}$ (df = 41; 306) \\ 
\hline 
\hline \\[-1.8ex] 
\textit{Note:}  & \multicolumn{1}{r}{$^{*}$p$<$0.1; $^{**}$p$<$0.05; $^{***}$p$<$0.01} \\ 
\end{tabular} 
\end{table}

\begin{table}[ht] \centering 
  \caption{Summary of the five most significant explanatory variables for predicting the minimum value of $B_z$ within the ICME.} 
  \label{tab3} 
\begin{tabular}{@{\extracolsep{5pt}}lc} 
\\[-1.8ex]\hline 
\hline \\[-1.8ex] 
 & \multicolumn{1}{c}{\textit{Dependent variable:}} \\ 
\cline{2-2} 
\\[-1.8ex] & Target \\ 
\hline \\[-1.8ex] 
 mean.bt. & 0.341$^{***}$ \\ 
  & (0.108) \\ 
  & \\ 
 std.bt. & $-$1.034$^{***}$ \\ 
  & (0.142) \\ 
  & \\ 
 cv.bt. & 0.522$^{***}$ \\ 
  & (0.094) \\ 
  & \\ 
 mean.np. & $-$0.645$^{***}$ \\ 
  & (0.079) \\ 
  & \\ 
 max.np. & 0.321$^{***}$ \\ 
  & (0.087) \\ 
  & \\ 
 Constant & $-$0.000 \\ 
  & (0.039) \\ 
  & \\ 
\hline \\[-1.8ex] 
Observations & 348 \\ 
R$^{2}$ & 0.473 \\ 
Adjusted R$^{2}$ & 0.466 \\ 
Residual Std. Error & 0.731 (df = 342) \\ 
F Statistic & 61.496$^{***}$ (df = 5; 342) \\ 
\hline 
\hline \\[-1.8ex] 
\textit{Note:}  & \multicolumn{1}{r}{$^{*}$p$<$0.1; $^{**}$p$<$0.05; $^{***}$p$<$0.01} \\ 
\end{tabular} 
\end{table}  

\begin{figure}[ht]
\centering
\includegraphics[width=5.0in]{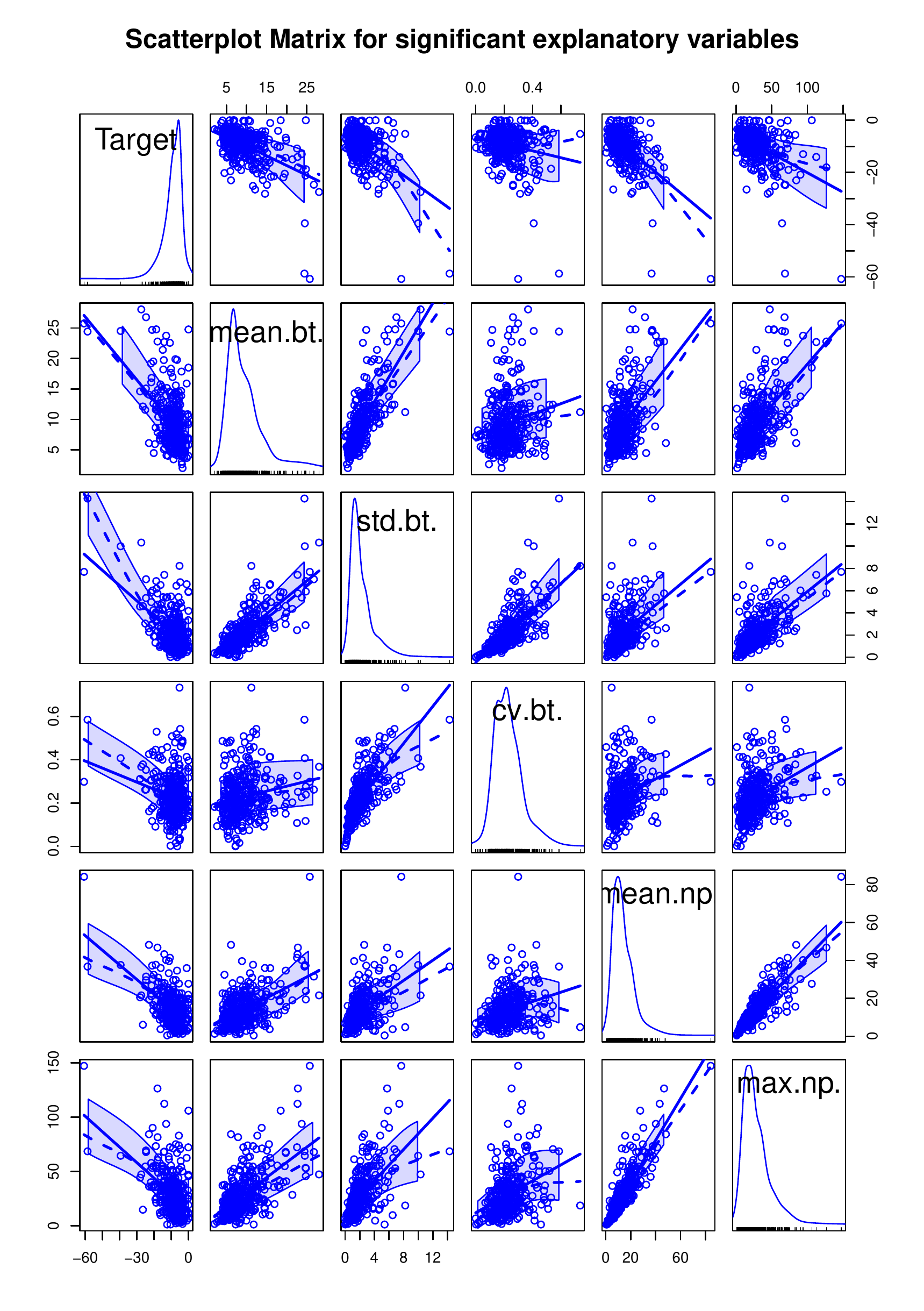}
\caption{{\bf Scatterplot matrix of five significant explanatory variables for the minimum value of $B_z$ within an ICME.}
Panels show: $B_z$ within the ICME (the output or ``target'' variable), mean.bt., std.bt., cv.bt.,  mean.np., and max.np. Data are shown by the circles, regression lines are solid, smoothed mean values are shown by the dashed line, and variances are shown by the dashed-dotted lines.}
\label{scatter}
\end{figure}

We can visually inspect the relationship between these statistically significant variables. Figure~\ref{scatter} shows a scatterplot matrix for these five most significant input (explanatory) variables and the target variable. Scanning down the first column or along the first row, we note that $B_z$ within the ICME (target) is negatively correlated with each of the five input variables. That is, larger negative $B_z$ values are associated with larger positive values of the input variables. However, we also note that the scatter is large, which is also reflected in the confidence regions. Additionally, the input variables are each positively correlated with one another, with the degree of correlation varying from one parameter to another, but, generally being higher within the $|B|$ or $n_p$ groupings. Comparing these groups results in a lower correlation. 

While the preceding analysis heuristically investigated the possible contribution of the explanatory variables in describing the variations in $B_z^{ICME}$, we next apply more robust techniques using a variety of algorithms to assess both statistical significance and importance. In these approaches, explanatory variables are iteratively added or removed to identify the subset of variables that produce the best-performing model, that is, the model with the lowest prediction error. 

First, using $R$'s MASS package, we applied both forward selection, where variables were added iteratively until the improvement is no longer statistically significant, and backward elimination, where variables are iteratively removed until the point is reached where all the variables are statistically significant (Table~\ref{tab4}). Based on this analysis, 22 variables account for 57\% of the variability in $B_z^{ICME}$. This is approximately the same as the 58\% that we computed using our {\it ad hoc} approach of searching through the variables. In this case, the most statistically significant parameters, in order of importance, were found to be: std.bt., mean.np., cv.bt., mean.bt., and max.np.

\begin{table}[!ht] \centering 
\tiny
  \caption{Step-wise regression using the MASS package.} 
  \label{tab4} 
\begin{tabular}{@{\extracolsep{5pt}}lc} 
\\[-1.8ex]\hline 
\hline \\[-1.8ex] 
 & \multicolumn{1}{c}{\textit{Dependent variable:}} \\ 
\cline{2-2} 
\\[-1.8ex] & Target \\ 
\hline \\[-1.8ex] 
 std.bx. & 0.117$^{*}$ \\ 
  & (0.065) \\ 
  & \\ 
 cv.bx. & $-$0.068$^{*}$ \\ 
  & (0.038) \\ 
  & \\ 
 min.by. & 0.247$^{***}$ \\ 
  & (0.076) \\ 
  & \\ 
 min.bz. & $-$0.195$^{**}$ \\ 
  & (0.078) \\ 
  & \\ 
 minmax.bz. & 0.058 \\ 
  & (0.038) \\ 
  & \\ 
 mean.bt. & 0.591$^{***}$ \\ 
  & (0.144) \\ 
  & \\ 
 max.bt. & $-$0.468$^{**}$ \\ 
  & (0.198) \\ 
  & \\ 
 std.bt. & $-$0.777$^{***}$ \\ 
  & (0.170) \\ 
  & \\ 
 cv.bt. & 0.596$^{***}$ \\ 
  & (0.098) \\ 
  & \\ 
 mean.vt. & 0.452$^{***}$ \\ 
  & (0.172) \\ 
  & \\ 
 std.vt. & $-$0.436$^{*}$ \\ 
  & (0.254) \\ 
  & \\ 
 cv.vt. & 0.472$^{**}$ \\ 
  & (0.221) \\ 
  & \\ 
 mean.np. & $-$0.642$^{***}$ \\ 
  & (0.147) \\ 
  & \\ 
 max.np. & 0.471$^{***}$ \\ 
  & (0.127) \\ 
  & \\ 
 min.np. & $-$0.171$^{**}$ \\ 
  & (0.081) \\ 
  & \\ 
 cv.np. & $-$0.154$^{**}$ \\ 
  & (0.070) \\ 
  & \\ 
 max.tp. & 0.257$^{**}$ \\ 
  & (0.115) \\ 
  & \\ 
 std.tp. & $-$0.335$^{***}$ \\ 
  & (0.123) \\ 
  & \\ 
 min.tp. & $-$0.144$^{**}$ \\ 
  & (0.058) \\ 
  & \\ 
 minmax.tp. & 0.143$^{**}$ \\ 
  & (0.056) \\ 
  & \\ 
 max.vt. & $-$0.255 \\ 
  & (0.168) \\ 
  & \\ 
 Constant & $-$0.000 \\ 
  & (0.036) \\ 
  & \\ 
\hline \\[-1.8ex] 
Observations & 348 \\ 
R$^{2}$ & 0.565 \\ 
Adjusted R$^{2}$ & 0.537 \\ 
Residual Std. Error & 0.680 (df = 326) \\ 
F Statistic & 20.179$^{***}$ (df = 21; 326) \\ 
\hline 
\hline \\[-1.8ex] 
\textit{Note:}  & \multicolumn{1}{r}{$^{*}$p$<$0.1; $^{**}$p$<$0.05; $^{***}$p$<$0.01} \\ 
\end{tabular} 
\end{table} 

Using the Random Forest method to assess the importance of the variables, we found the eight most important variables (in order of importance) were: mean.np., min.np., min.by., std.by., max.bt., mean.tp, max.bz., and mean.bt (Table~\ref{tab5}). Beyond this, there was a drop in relative importance. We note that bz-based variables are likely to be associated with $|B|$ within compression regions, thus, to some extent they can be considered confounding variables. 
Comparing this list with the significant variables identified earlier suggests that any subset of significant {\it and} important explanatory variables would include -- at least -- include: mean.bt., max.bt., and mean.np. It should be noted that re-running the Random Forest Method repeatedly resulted in different estimates - this is a statistical method employing random variables. Thus, particularly as we step down through the variables, less confidence can be given to their relative positions. For example, cv.tp. and minmax.tp., for example, often change their relative strength, although their ratios with respect to mean.np, for example, always remains significantly less than one. 

\begin{table}[!ht] \centering 
  \caption{Random Forest Method for the relative importance of parameters.} 
  \label{tab5} 
\begin{tabular}{@{\extracolsep{5pt}}lc} 
\hline 
Parameter & Overall \\
\hline
              Overall \\
max.bx.    & -0.2629452 \\
std.bx. &     1.3208674 \\
min.bx.   &   1.4478551 \\
cv.bx.  &     2.4411125 \\
minmax.bx. & -2.5282641 \\
mean.by.    & 3.3642488 \\
max.by.     & 5.4092380 \\
std.by.     & 8.0053151 \\
min.by.     & 8.5938032 \\
cv.by.     & -1.6325343 \\
minmax.by.  & 1.3254579 \\
mean.bz.    & 1.0951099 \\
max.bz.     & 7.4562721 \\
std.bz.     & 5.8829339 \\
min.bz.     & 3.0405328 \\
cv.bz.      & 0.2845630 \\
minmax.bz.  & 1.9436885 \\
mean.bt.    & 7.1670919 \\
max.bt.     & 7.7885147 \\
std.bt.     & 4.0623384 \\
min.bt.    & -2.4643346 \\
cv.bt.     & -0.9527299 \\
minmax.bt.  & 1.6490376 \\
mean.vt.    & 4.3299150 \\
max.vt.     & 4.1455184 \\
std.vt.     & 1.3582930 \\
min.vt.     & 1.7991820 \\
cv.vt.      & 1.6273622 \\
minmax.vt.  & 2.8107001 \\
mean.np.   & 12.5156175 \\
max.np.     & 5.8775704 \\
std.np.     & 1.2936570 \\
min.np.    & 11.3617008 \\
cv.np.      & 1.9204835 \\
minmax.np.  & 2.7836057 \\
mean.tp.    & 7.5204233 \\
max.tp.     & 2.9264847 \\
std.tp.     & 3.8501390 \\
min.tp.     & 4.2046862 \\
cv.tp.      & 3.6325805 \\
minmax.tp.  & 2.9420327 \\
\hline 
\end{tabular} 
\end{table}

Applying the Xgboost method for ranking explanatory variables in order of their importance identified min.by., max.bt., mean.np., mean.bt., std.by., min.np., mean.by., and mean.tp. (Table~\ref{tab6}). Beyond these, the remaining variables were only 10\% or less of the power of the most important variable. Again, we note that features associated with $|B|$, $n_p$ and $B_y$ dominate the list. 

\begin{table}[ht] \centering 
  \caption{Xgboost Method for relative importance of parameters. Top-20 shown.} 
  \label{tab6} 
\begin{tabular}{@{\extracolsep{5pt}}lc} 
\hline 
Parameter & Overall \\
\hline
min.by.    & 100.000 \\
max.bt.     & 88.261 \\
mean.np.    & 32.384 \\
mean.bt.    & 23.530 \\
std.by.     & 18.147 \\
min.np.     & 17.399 \\
mean.by.    & 14.750 \\
mean.tp.    & 13.012 \\
max.bx.     & 10.333 \\
minmax.np.   & 6.411 \\
cv.vt.       & 5.980 \\
max.bz.      & 5.911 \\
cv.bt.       & 5.464 \\
cv.bx.       & 5.017 \\
minmax.tp.   & 4.914 \\
cv.by.       & 4.499 \\
max.tp.      & 4.311 \\
min.tp.      & 3.853 \\
minmax.bz.   & 3.809 \\
minmax.bx.   & 3.754 \\
\hline 
\end{tabular} 
\end{table} 

The Multivariate Adaptive Regression Splines  (MARS) model can also be used to rank explanatory variables. It is a flexible technique and is included here to provide evidence for the sensitivity of our results to the technique implemented (Table~\ref{tab7}). 
Thus, we remark that cv.vt and std.vt are given a higher relative position than mean.tp., say, which is in contrast to the random forest (Table~\ref{tab5}) and Xgboost (Table~\ref{tab6}) methodologies. 
Again, variations in $B_z^{ICME}$ are explained primarily by features in $B_y$, $n_p$, and $|B|$. 

\begin{table}[ht] \centering 
  \caption{MARS Method for the relative importance of parameters. Results are sorted by order of descending importance given by measures gcv and rss.} 
  \label{tab7} 
\begin{tabular}{@{\extracolsep{5pt}}lccc} 
\hline 
Parameter & nsubsets  & gcv &   rss \\
\hline
min.by.        & 12 & 100.0  & 100.0 \\
mean.np.       & 11  & 59.8   & 63.1 \\
mean.bt.       & 10  & 49.8   & 53.8 \\
min.np.         & 8  & 36.1   & 40.9 \\
cv.vt.          & 7  & 26.9   & 33.0 \\
std.vt.         & 5  & 18.3   & 24.7 \\
max.np.         & 5  & 18.3   & 24.7 \\
mean.tp.        & 4  & 17.9   & 23.2 \\
std.np.         & 4  & 17.1   & 22.5 \\
\hline 
\end{tabular} 
\end{table} 

\begin{figure}[ht]
\centering
\includegraphics[width=4in]{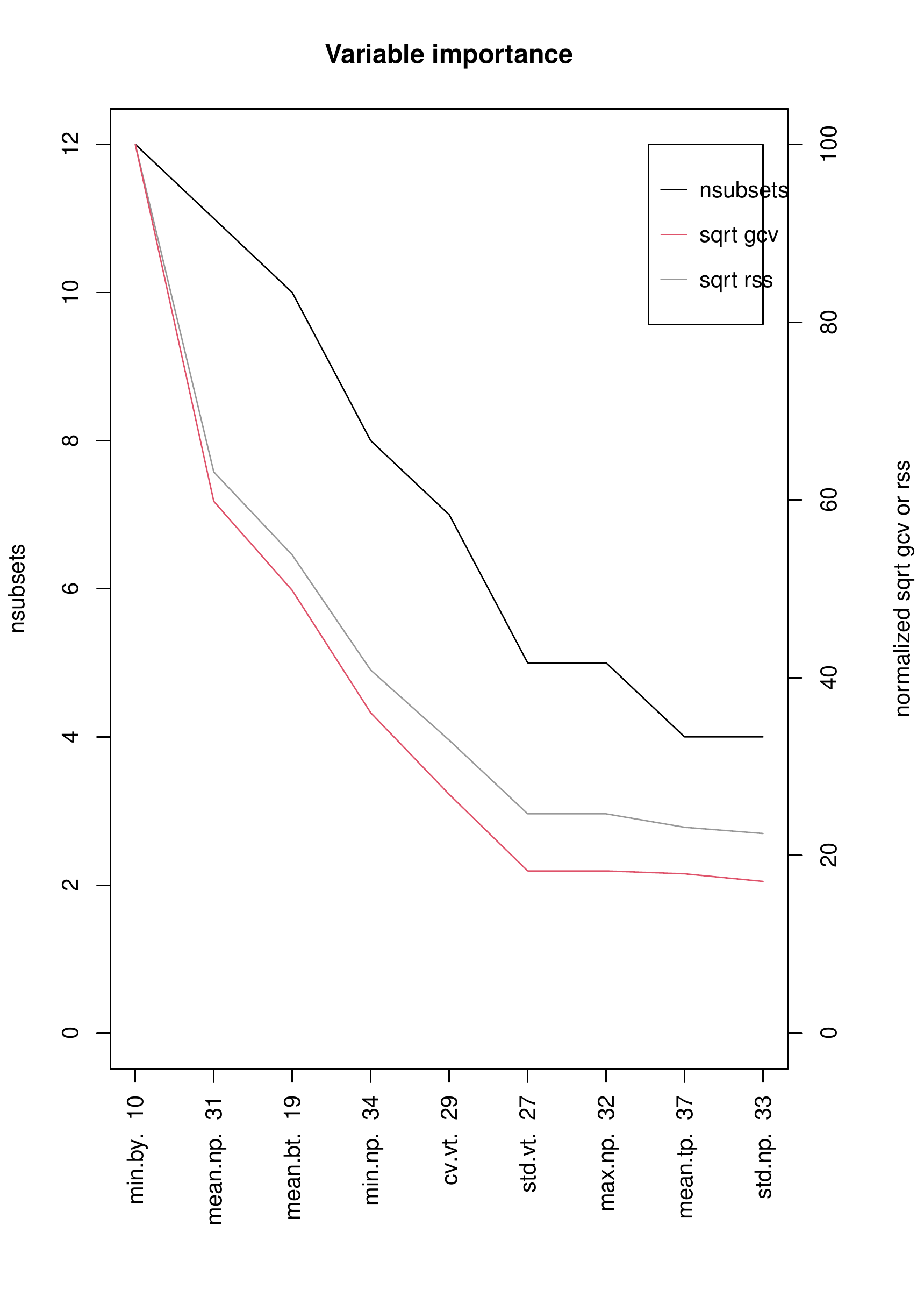}
\caption{{\bf MARS Method for the relative importance of parameters.}
See text for more details.}
\label{mars}
\end{figure}

We can also explore these results graphically (Figure~\ref{mars}), which highlights the relative contribution of the variables. In particular, we note that, in this case, the first four variables account for most of the variability in $B_z^{ICME}$.

The Stepwise regression method can also be combined with the Akaike Information criteria (AIC) to identify the best model, that is, the best combination of parameters to explain the output variable. 
Higher values of AIC provide more support for a particular variable's importance. 
Using this approach, in order of importance, mean.np., cv.bt., and std.bt. were all found to be particularly important (Table~\ref{tab8}). 

\begin{table}[ht] \centering 
  \caption{Step-wise Regression Method for the relative importance of parameters. Only the 10 most important parameters are shown in increasing order of importance.} 
  \label{tab8} 
\begin{tabular}{@{\extracolsep{5pt}}lcccc} 
\hline 
Parameter   &    Df  & Sum of Sq   &  RSS  &    AIC \\
\hline 
none   & $...$ &  $ ... $         &        156.29 & -246.56 \\
... & ... & ... & ... & ...  \\

cv.np.     &   1   &   1.8425 &  158.13 &  -244.49 \\
max.bz.    &   1   &   2.6053 &  158.90 &  -242.81 \\
std.bz.    &   1   &   3.6149 &  159.91 &  -240.61 \\
mean.bt.   &   1   &   4.1432 &  160.44 &  -239.46 \\
max.np.    &   1   &   5.2433 &  161.53 &  -237.08 \\
mean.tp.   &   1   &   6.4323 &  162.72 &  -234.53 \\
min.by.    &   1   &   6.4458 &  162.74 &  -234.50 \\
mean.np.   &   1   &   9.2595 &  165.55 &  -228.54 \\
cv.bt.     &   1   &  15.5157 &  171.81 &  -215.63 \\
std.bt.    &   1   &  15.7451 &  172.04 &  -215.16 \\
\hline 
\end{tabular} 
\end{table} 

\begin{figure}[ht]
\includegraphics[width=\columnwidth]{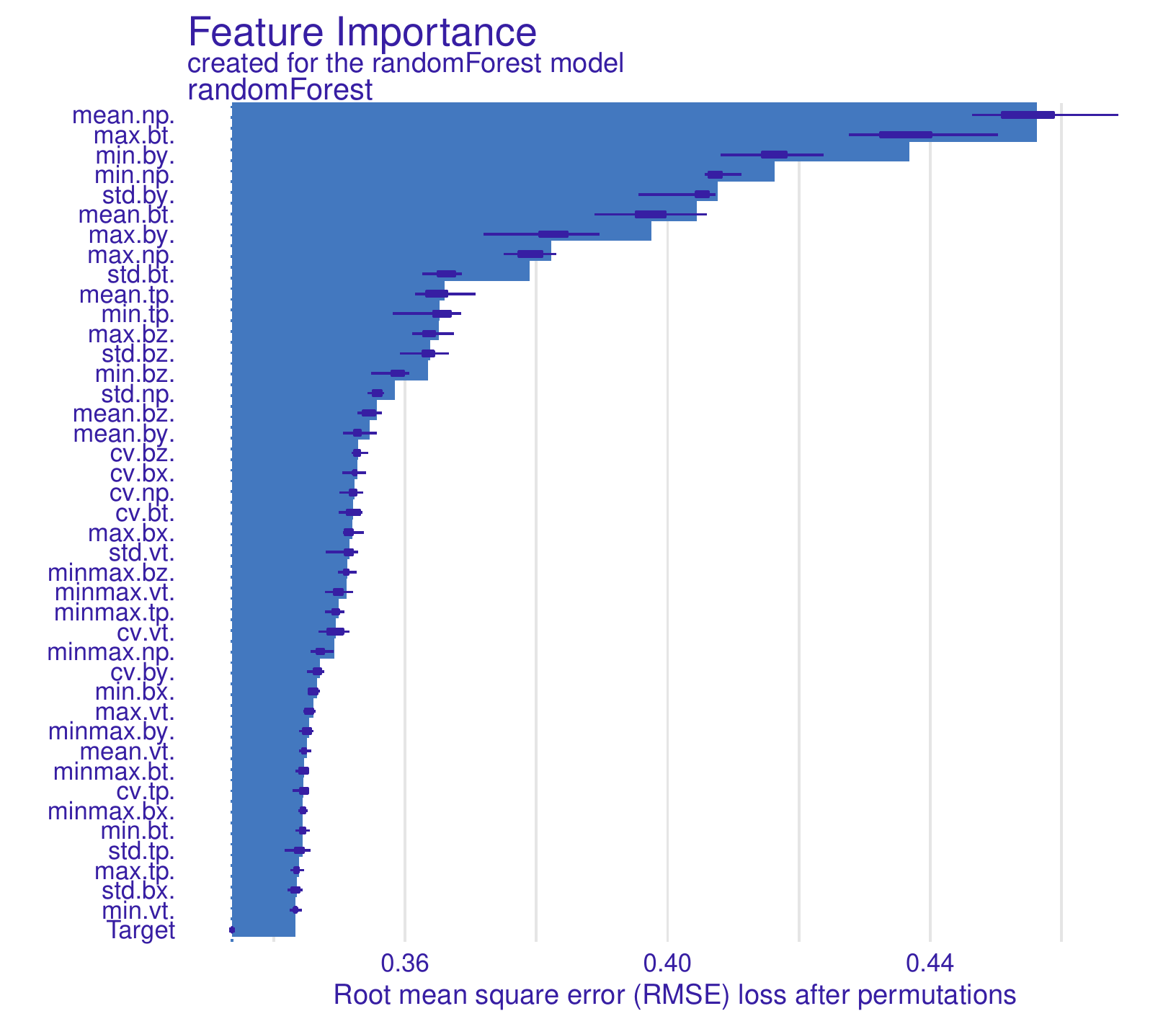}
\caption{{\bf Relative importance of explanatory variables using the DALEX method. }
The RMSE loss is a measure of the relative importance of a particular input variable.}
\label{dalex}
\end{figure}

Finally, we consider the DALEX package, which is, in fact, a meta-package in the sense that it can compare responses from different models to allow for direct comparison. Here, though, we apply the DALEX machinery with the random forest technique and use the `explain' and `variable\_importance' functions to quantify their relative importance (Table~\ref{tab9}). In this case, mean.np., max.bt., min.by., std.by., min.np., and mean.bt. describe most of the observed variability in $B_z^{ICME}$, with a large jump between mean.bt. and the next variable, max.np. The degree to which these variables contribute to the overall variations can also be visualized graphically (Figure~\ref{dalex}). From this, we infer that the first six variables account for most of the observations.  

\begin{table}[ht] \centering 
  \caption{DALEX Step-wise Regression Method for relative importance of parameters.} 
  \label{tab9} 
\begin{tabular}{@{\extracolsep{5pt}}lcccc} 
\hline 
 No.      &        variable & mean dropout loss    &    label \\
\hline 
1  & full model         & 0.3336585 & randomForest \\
2  &       Target         & 0.3336585 & randomForest \\
... & ... & ... & ... \\
35   &    std.bt.         & 0.3660815 & randomForest \\
36   &    max.np.         & 0.3789919 & randomForest \\
37   &    max.by.         & 0.3822500 & randomForest \\
38   &   mean.bt.         & 0.3975788 & randomForest \\
39   &    std.by.         & 0.4044210 & randomForest \\
40   &    min.np.         & 0.4075862 & randomForest \\
41   &    min.by.         & 0.4163421 & randomForest \\
42   &    max.bt.         & 0.4368321 & randomForest \\
43   &   mean.np.         & 0.4562865 & randomForest \\
44   & baseline         & 1.2421413 & randomForest \\
\hline 
\end{tabular}
\end{table}

\subsection{Re-Analysis using Updated Database}

Although we focused on the same dataset reported by \citeA{reiss21a}, we repeated the analysis with an updated ICME dataset (v2.1)  containing a total of 521 events. We found that the results were broadly consistent with the earlier analysis, although, in general, the associations were found to be slightly worse, likely the consequence of the more recent CMEs being generally weaker. Thus, while the main explanatory variables remained the same (e.g., those derived from np and bt), those of lower importance (but still statistically significant) were redistributed modestly within each table. 

Finally, we repeated the analysis, this time separating out the events into those that were preceded by shocks and those that were not. The results were generally consistent with our expectations that the relationships were strongest for the group that contained shocks and weakest for those that did not. However, the effect was modest. For example, the adjusted-$R^2$, a measure of model accuracy that attempts to mitigate the effects of over-fitting, or in our case, address the differences in the sample size amongst the datasets, was found to be:  0.500, 0.524, and 0.542 for the no-shock, all-events, and shock-only datasets. Thus, the adjusted-$R^2$ increases as the proportion of shocks in the dataset increase, which we would intuitively expect, but also, the differences between all three datasets are relatively modest. 

\section{Summary and Discussion}

In this study, we found that statistical measures tracking number density and magnetic field strength accounted for a large proportion of the variability of the large-scale z-component of the magnetic field within ICMEs. Taken together, these features capture the degree to which the ICME compresses the ambient solar wind ahead. Intuitively, this makes sense: Energy made available to CMEs as they erupt is partitioned into magnetic energy density and kinetic energy. Thus, more powerful CMEs are launched with larger flux-rope fields (larger $B_z$), and at greater speeds. Faster CMEs produce greater sheath compression leading to increased number densities and total field strength ahead of the ICME. 

In some methodologies, other potentially relevant features were found, such as the mean proton temperature within the sheath region. This too is related to the compression of the region since the plasma heats up as it is squeezed. The fact that temperature was not found to be consistently as important in explaining the field variations within the ejecta may be related to the intrinsic noise associated with these measurements, thus, reducing the signal component. It is also worth pointing out that features such as those related to $v_x$ and $B_x$, which would not be modified when compressed, were notably absent (or relegated to lower importance) in most of the results, strengthening the idea that compression is the key mechanism connecting the features to the targets. 

In our study, whether the sheath region was bounded by a shock did not appear to make a significant difference. Hence, this approach works for events that are not sufficiently strong as to drive an observable shock. All that is required is a sheath region. On the other hand, the association was stronger for the events with shocks. Thus, we suggest that this approach is likely more accurate for these events. Additionally, adding more recent events to the dataset modestly reduced the associations. This suggests, at least tentatively, that there may be a solar cycle effect, although more work would need to be performed to support or refute the idea. 

It should be emphasized that these results are statistical in nature. There is always a danger of over-interpreting the results or ascribing a causal relationship to a seemingly important feature that turns out to be spurious. Those associated with $B_y$, are a good example. What is the significance of min.by? It does not appear to be directly related to compression, in the more obvious way that mean.by or max.by might be. We could conceive of a mechanism whereby negative values of By rather than positive values (i.e., large minima) might be preferentially associated with one magnetic sector (inward) over another, but since the data span an entire solar cycle, it is not clear that such an asymmetry would exist. Moreover, even if it did, it is not clear what the mechanism would be. Instead, it may just be an indirect measure of the overall compression of the magnetic field, and more visible because it has less noise associated with it than, say, max.by. Since its relative importance shifts depending on the ML approach used, we must also recognise that its presence may be no more than a statistical anomaly. As noted previously, different techniques rely on different objective functions, thus, the optimum prioritised list of features would be expected to change modestly from one technique to another, and, certainly, as one steps further down in importance. 

Our results can be compared with more traditional approaches for connecting upstream to downstream properties of solar wind structures. \citeA{borovsky10a}, for example, used a superposed-epoch analysis of corotating interaction regions to quantify how solar wind parameters change in crossing stream interfaces. More recently, \citeA{regnault20a} used a similar technique to reveal some generic features of ICME profiles, finding that fast ICMEs still retained signs of compression at 1 au, and consistent with our results. Additionally, they were able to reconstruct a general asymmetric profile in the magnetic field, which was not possible with our approach. Finally, \citeA{salman20a} analysed the properties of ICMEs and their associated sheaths using both statistical and superposed-epoch approaches. They separated their events into: (1) CMEs with no sheath; (2) CMEs with a sheath but no shock; and (3) CMEs with both a sheath and a shock, finding that the two latter categories correlated well with the speed of the CME. In summary, they found that sheaths driving shocks tended to be more compressed (higher densities and field strengths) and hotter. 

This study could form the motivation for several follow-on investigations. For example, using the updated v2.1 CME catalogue, the dataset could be used to pre-select those events that are likely to be most geoeffective. These would likely be the fastest ICMEs, driving the strongest sheath regions. We anticipate that the associations would be even stronger for these events. Additionally, other estimates of the compressive properties of ICME sheaths, such as from heliospheric imager or IPS measurements, could be used to estimate the likely z-component of the magnetic field within the ICME, which could, in principle, lead to longer forecasting lead times. 

Our results expose the inherent sensitivity and limitations in applying complex statistical analyses to datasets containing both noise and a signal that is built up from complex physical processes. While these techniques potentially reveal the real relative importance of the features, it may not always be possible to separate them from statistical fluctuations driven by noise within the data. This is underscored by the re-analyses of the data using different renormalisations of the data. While this did not materially affect the main results of the study, it did re-shuffle the relative importance of some of the more minor, yet statistically significant features. In spite of these limitations, we suggest that our approach of applying many techniques to the same dataset is a valuable exercise, in that it allows us to assess the likelihood of over-interpreting the results by focusing solely on one approach. 

In closing, we reiterate the main points of this study: (1) ML techniques can be used to infer the z-component of the magnetic field within ICMEs using readily available measurements from in situ spacecraft upstream of them; (2) the most important and statistically significant features are related to the sheath plasma density and field strength; and (3) the physical interpretation of this result is intuitive -- these parameters capture the degree of compression upstream of the CME, which correlates with the ICME's speed, and, in turn, the overall strength of the magnetic field within the ejecta.

\acknowledgments
PR gratefully acknowledges support from NASA (80NSSC18K0100, NNX16AG86G, 80NSSC18K1129, 80NSSC18K0101, 80NSSC20K1285, 80NSSC18K1201, and NNN06AA01C), NOAA (NA18NWS4680081), and the U.S. Air Force (FA9550-15-C-0001). MAR and CM were funded, in part, by the European Union (ERC, HELIO4CAST, 101042188). Views and opinions expressed are however those of the author(s) only and do not necessarily reflect those of the European Union or the European Research Council Executive Agency. Neither the European Union nor the granting authority can be held responsible for them.”

\section*{Open Research}

\subsection*{Data Availability Statement}


All data and code used in this study can be accessed from the Zenodo \cite{riley23a}. 


%
%


%
%
%
%
%

\end{document}